# DRAFT
# Procedural animations in interactive art experiences -
# A state of the art review

C. Tollola
ctoll103@uottawa.ca
University of Ottawa

**1. Abstract**

The state of the art review broadly oversees the use of novel research utilized in the creation of virtual environments applied in interactive art experiences, with a specific focus on the application of procedural animation in spatially augmented reality (SAR) exhibitions. These art exhibitions frequently combine sensory displays that appeal, replace, and augment the visual, auditory and touch or haptic senses. We analyze and break down art-technology related innovations in the last three years, and thoroughly identify the most recent and vibrant applications of interactive art experiences in the review of numerous installation applications, studies, and events. Display mediums such as virtual reality, augmented reality, mixed reality, and robotics are overviewed in the context of art experiences such as visual art museums, park or historic site tours, live concerts, and theatre. We explore research and extrapolate how recent innovations can lead to different applications that will be seen in the future.

**2. Introduction**

Interactive art is an exciting field for researchers, technologists and artists, leveraging modern research to provide art entertainment to the masses. In 2020, the need for social distancing during a global pandemic has increased demand for experiences that do not require travel, or are human-made. Procedural animation refers to the generation of rules-based motion generation, stemming from inverse and forward kinematics. These rules generally apply to robotics control, but computer software such as open-sourced Unity and Maya allow amateurs, established production studios, programmers and artists to benefit from a suite of tools to create realistic animations that apply these formulas in several fields such as virtual reality, computer games, and movies. Animation is the essential ingredient for a visually-focused interactive art experience, it is also graphics-heavy, TeamLab's "Borderless" digital art museum, synchronically renders animations in 8K resolution and laser projectors real-time, providing visitors to the museum a dark walk into a mixed and augmented reality without the use of head-mounted displays.

Several advances in spatial augmented reality attempt to solve latency of media, calibration, accuracy presenting depth-of-field, 3D reconstruction of real buildings and objects. Therefore, we present the essential innovations in calibration, resolution, the unique artistic effects that propagate and exist due to both hardware and algorithms developed and used to construct and fabricate technology in suggestively its most direct, and artistic, form. Through the exploration of limitations of such various technologies in interactive art, conclusions are drawn, how do we create an experience to connect humans to technology, and inspire them to believe the world around them has changed from reality, applying this concept to inspire future generations to innovate through technology.

We present the following topics:
- Case studies and emerging applications
- Projection mapping
- Hardware
- Possible artistic effects and animations

## 3. Case Studies and Emerging applications

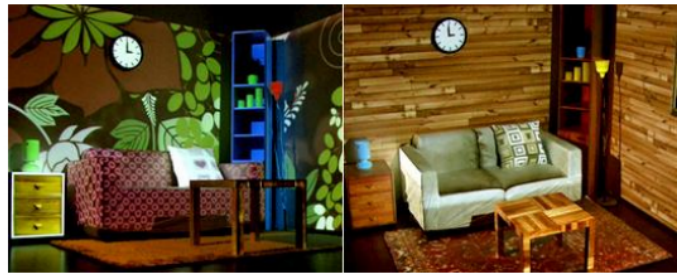

Figure 1: A theatrical stage (right) enhanced by 3D projection mapping (left)

In recent years, Virtual Reality (VR), Augmented Reality (AR), Mixed Reality (MR), Extended Reality (XR), Projection Mapping/ Spatial Augmented Reality (SAR) and robotics have become considered the most technologically advanced artistic mediums. These mixed technologies extend and have the ability to display and computerize traditional artistic disciplines such as visual art, film, games, music, theatre, and puppetry. The depth of man-made reality changes according to which of the introduced display mediums are applied. In the mid-1990s, these terms were conceptualized as the "virtuality continuum" [1] that described the spectrum of mediums to be "entirely real" to "entirely virtual".

VR involves computer-generated visuals and environments that completely surround the real-world of a user. The user has a central view and is able to view and/ or interact with the virtual environment in real-time. Highly engaging hardware such as a VR head both visually and auditorily immerses the user.

Mixed reality is a broad term that blends digitally created content and a real-world environment which can involve AR and VR simultaneously. MR can encapsulate or be compared to immersive entertainment/ hyper-reality which involves SAR or projection mapping virtual content onto real environments. Generally, SAR involves camera-projector systems capable of projecting laser displays onto flat or 3D surfaces. XR is an umbrella term and used to group VR, AR and MR together and software, hardware, methods and experience to see these levels of virtual reality.

The fluidity and realism of interaction between the virtual art experience and a participant vary based on the devices chosen explained in section 5.

In this section, we introduce applications that attempt to transport users into spaces that are completely mad-made. Evidently, when computers make it possible to virtually travel to distant environments, "here" and "now" may begin to take on new meanings.

## 3.1 MR, XR and Projection Mapping (SAR) Applications

Hyper-reality applications have been tremendously present in recent years due to technological advances. The concepts have been applied to physical environments: museums, amusement parks [c1], arcades, and concerts [7,c2]. This section also focuses on projector-camera systems where various objects, such as an object held by the user, clothes, a human body, and a face, are projection targets [26].

Laila [4] is an opera experience with a storyline produced by the Finnish national opera, in the space of a 360-degree dome. In August, 2020 it will attempt to place users as the driving forces for animations displayed by laser projectors and partially generated by artificial intelligence. Procedural animations such as flocks of birds are affected by parameters of swarm intelligent algorithms and chaos theory. Triggers by the audience affect the rendering of bird motivations, including their group separation sensitivity, ease of alignment, and cohesive movement. They otherwise act according to field functions such as: it is easier to go in one direction than another. Examples of triggers are user movement and group dynamics in the space which affect the birds that fly in flocks on the screen. The butterfly effect implemented introduces unpredictability in animations, such that small initial changes in the system created by interaction may accumulate into larger changes later on. The specific hardware implemented in the project is not known.

In 2018, mixed reality creation studio TeamLab opened "Borderless" in Tokyo It is a dedicated 10 000 square meter projection mapping museum where corridors lead to different spaces in which a different artwork of a different style is displayed, generally in the form of interactive animations. In 2019, it drew 2.3 million visitors from 160 countries around the world. Some 520 computers, 470 Epson projectors of 6,000 to 15,000 ANSI lumens are used to render real-time animations in 8K quality [5]. The museum has shown different artwork over the years. In one "interactive display", users are able to touch flowers projected on walls and a real-time render dissipates the flower [video 1]. Such animations use generated animation randomness for unique content similarly to Lilia but are unknown. Specific sensors are not known, but could involve depth sensors, RGB cameras and LiDAR. Localised and spatial sound in soundscapes add to immersion and transition between corridors and into the experience by Yamaha VXC and VXS series loudspeakers. A tearoom area in the museum leverages pose estimation and object tracking to project different animations onto tea in a cup or tea spilled [video 2].

In 2019, art installation "gravityZERO" combined a programmed robotic arm to suspend a participant, projection mapping for projection of realistic images of outer space, and music to mimic the sensation of zero-gravity [v].

A projection mapped gym with virtual targets was developed to allow sports interaction between players with mobility issues and those without [19].

Recently, jelly type edible retroreflectors have been applied directly on a pancake [24] to be detected by projection mapping systems. They can be detected in low light conditions, an advantage over edible QR codes [25]. This is another medium on which to project art and possibly interact with.

Concerts streamed in real time over the internet in 2020 have evolved into a mainstream phenomenon, with a single concert grossing about $20 million in sales [6]. Other such "livestream" adoptions include an artist streaming a concert in a massive multiplayer online game virtual environment [7], a television singing competition demonstrating projection onto real-life objects and set, [8]. Other applications include dance [14] and a concert using AR animations [9]. Projection mapping in livestream concerts followed shortly after [video 3]. Popular social media platform Facebook will introduce a paid streaming platform [10], due to high demand for live interactive art experiences.

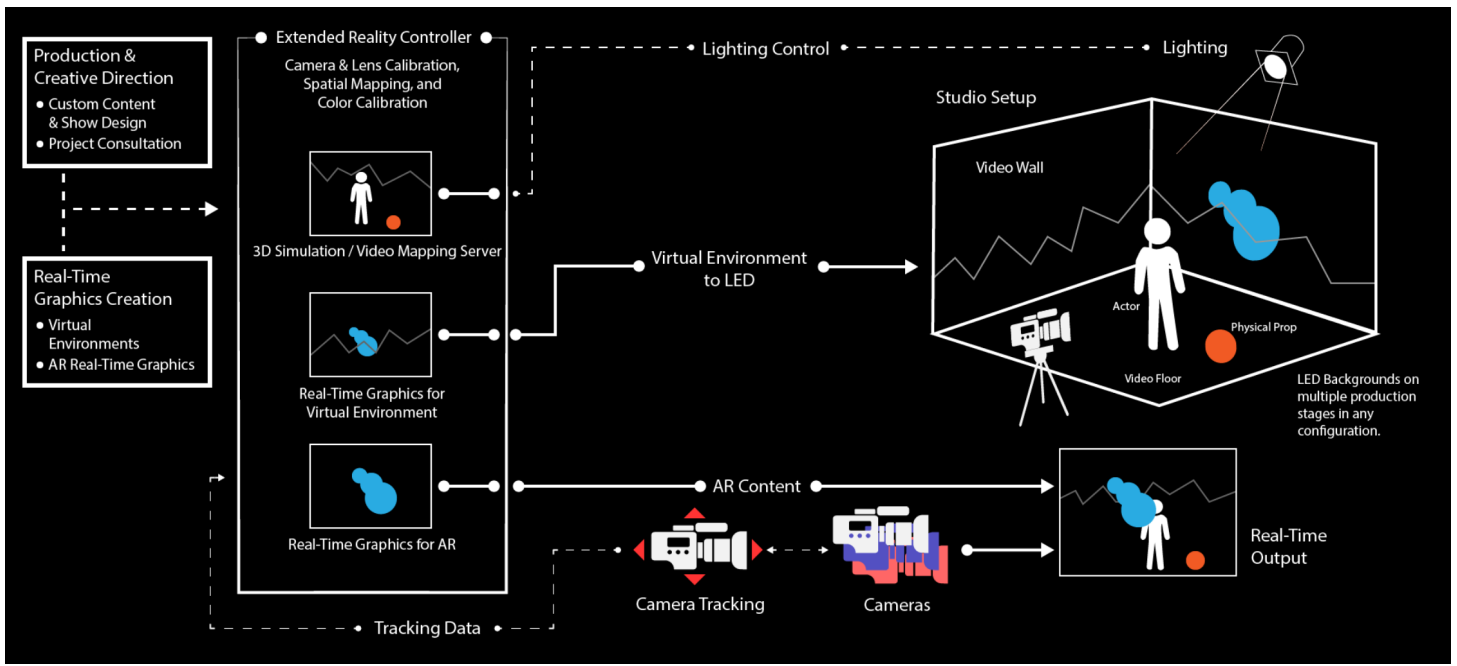

Figure 2: [8] Application of 3D graphics simulation in a singing performance of background LED panel uses tracks the camera movement to render a realistic 3D background

### 3.2 AR Applications

Art and graphics applied to augmented reality have shown to enhance application areas such as sports, zoos, historic sites, and could see application in the fashion industry.

Zoo [33] where users wearing a Magic Leap One headset can view the names and details of different zoo animals, view AR skeletal structures applied to the animals, play mini games with generated images as well as use an AR generated tape measuring tool. The AR zoo experience involves the surrounding spatial 3D environment, controller touch and triggers, 3D audio elements and head tracking. 360 video

Panasonic demonstrated how AR feedback in ping pong allows the audience to feel immersed during a match [11]. Museums and zoos have used AR to annotate and provide commentary on exhibits. One such project was developed for the Singapore cinema is also exploring annotation [27] through AR and an agenda has been proposed for television watching [28] .

AR is proposed to revitalize cultural heritage sites. In Brisbane, UK an AR adventure game is proposed to deliver a historically accurate and sense of "spirit" to recreate a tour experience [u]. The heritage site only previously had signage to communicate the history of the house to visitors. AR art attached to a physical object is being sold with capabilities to connect to AR headsets [13].

In 2018, Disney successfully created a system that overlays a person with a "watertight" 3D costume, using only a monocular RGB image [q]. They use a 3D costume model, manually create a 3D skeleton with different proportions and poses wearing the costume, and finally best-match the resulting 3D costume data set to the 2D participants' skeleton taken from 2D pose tracking.

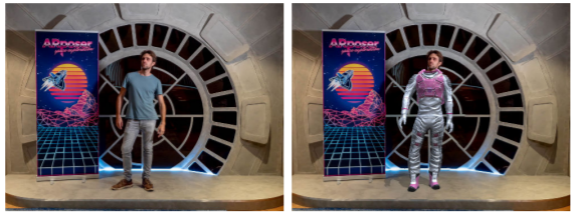

Figure 3: 3D costume augmentation from a single monocular camera image

One limitation of the work is that greater accuracy of overlaying the costume relied on fine tuning parameters to certain people (ie: limb length variance) or poses. Another limitation of the method is that it lacks robust pose and shape estimation which could be solved by machine learning optimized to run in real-time. This innovation could be extended to work in art experiences such as a means of trying on clothing, either for historic museums, or fashion runway shows which have struggled with interactivity during a global pandemic [m] since AR has already been used in fashion magazines [e].

Virtual puppeteering is another AR implementation where a user can use their smartphone to move and interact with different animated characters through their camera. The system leverages procedural animation such that predefined motions for a given character are "rigged" with joints using Unity and interactions with the user trigger generated motions. The puppeteering application uses a state machine approach to determining rules to animated character movement, such as a state is defined by internal logic, the root position of the character, the configuration of the animation, and the inverse kinematic state.

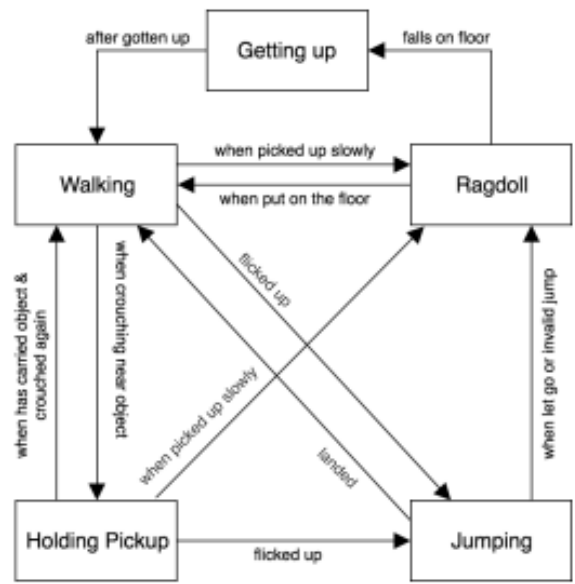

Figure 4: State machine used between puppet character animations

### 3.3 VR Applications

Virtual reality is the highest level of immersion currently offered by research, and generally involves a head-mounted display that may include headphones and controllers. A participant may use virtual reality to explore a new environment, watch a movie, or play a game. The viewing of 360 degree video and image artworks in fine art, film, and photography disciplines that are compatible with headsets, but do not require them, have been debated as virtual realities themselves due to their 3D-mimicking nature.

An example of virtual reality applied in 2020 is a World War II VR experience that combines realistic sound and historical data of real war trenches to deliver an immersive educational experience [29]. Major publishing company National Geographic also published a cinematic VR series from the point of view of real soldiers [30].

Another VR experience takes the form of a virtual impressionist-style painting world. Visitors to the world become brush strokes themselves as they move in space [31].

A VR pottery workshop experience received high appraise in 2019, allowing a new virtual medium for art creation [t]. As well, a fully VR museum was created with a popular block-shaped virtual world builder Minecraft with areas for visitors to create their own art and disrupt museum exhibits [32]. In video games, the use of a live host in multiplayer escape games has been proposed to improve interactivity between players [34].

Extending this idea of interactivity as art; social art experiences through games have increased - VR social experience RecRoom launched in 2016 now receives 40 million monthly visits. It includes 3 million user-created games created using an in-game visual node-based programming language [35]. Furthermore, social media platform Facebook is setting the stage for mainstream adaptation of VR by launching a similar platform currently in beta release, Horizons [36].
A more defined example of a social VR game is a collaborative baseball batting cage intersects a social art and sports experience.

In the film industry, film festivals have accepted VR films since 2016 [38] Over 15 VR and XR film festivals have taken place online in 2020 [12]. A notable example of interactive film executes an interactive narrative where the watcher can make decisions to effect the ending [39, s].

The immersive filmmaking studio, Felix & Paul released a VR series "Space Explorers: The ISS Experience", in 2018 for Oculus headsets and 360 video enabled mobile devices that depicts the lives of NASA astronauts on the International Space Station (ISS). The last episode will film the first-ever spacewalk in VR outside of the ISS. The studio modified a Z-Cam V1 Procamera with nine 4K sensors to create a 8K 360-degree [e].

### 3.4 Robotics and sensing technologies

Social robots have been applied for the purpose of interactive storytelling for young children. A robot made as a companion for children with cancer invited children to make choices about the story, to reenact parts in the story with it together, and to record and replay different sound effects. Compared to conventional media, the study concluded that the robot improved enjoyment of the story, and that children recalled more about the story [40]. Robots in standup comedy have been improved to tell jokes with accurate timing [41], receiving 69.5% positive reactions to jokes from a live audience and therefore could see realistic employment in the future.

Aerial robots - drones, have grown as an artistic medium due to the introduction of swarm path-planning. In 2017 their GPS positioning abilities were used at a Superbowl in America to display the US flag with mounted LED-panels [42]. In 2018, an artist used a drone to create graffiti artwork [43]. Disney also developed an autonomous spray-painting drone [r]. Through various path-planning algorithms, drones have produced shows similar to fireworks and these algorithms have been compared in [p]. The government of South Korea used 300 drones to display an image depicting essential workers using masks during the Covid-19 pandemic [b]. In the future an application might be created to allow artists without the expert knowledge in path-planning and drone limitations to choreograph their own drone shows. Drones have also been used as haptic devices for VR/ AR outlined in the haprics section. A dedicated museum to drone art and drones-as-sculptures in Lisbon has been investigated [44].

The idea of utilizing a network of sensors to recreate a real marsh into a virtual reality has been applied [45] by the MIT Media Lab and NYU with "Doppelmarsh". The virtual reality uses textures extracted by camera from the real environment to be rendered as "Doppelmarsh" visual assets using the Unity3D game engine. Furthermore, wind, temperature, precipitation sensors as well as cameras to detect rain, snow, fog and changes in season are sent to a websocket server to facilitate sensor-data mapping which is then analyzed with Google Vision API, stored, and rendered by the game engine

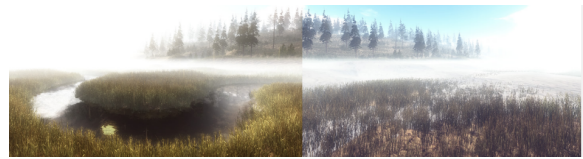

Figure 5: The virtual marsh and the real world marsh

The Doppelmarsh's "reality resynthesis" can be applied to a variety of mediums such as VR and be used as a means to experience real-life locations. For example, artists created the outer space themed installation "Orbits" that follow the movements of space junk, espionage satellites, and normal satellites in real-time, The installation also produces a complex soundscape of sine waves whose frequency depend on the altitude of a specific satellite above the surface of the Earth and volume depends on its distance from the observer [c]. Another sensor musical installation produces a composition dependent on live data of sharks tracked with GPS.

### 3.5 Summary

Through this section, we have shown that emerging technologies are often applied in art as an experimental vehicle. Other applications of interactive art do not fit into any of the categories beforehand.
The first is "video art" which utilized generative neural networks to photorealistically replace the face of an actor in the video [g]. An updated version of this recent media phenomenon is explained in section 6. The second is a co-created art experience where professional artists received users on a cloud-hosted video platform, and adjusted the participants' video feed stylistically [y]. The application of and mix of extended reality, and robotics technology will only increase in the field of interactive art.

### 4.1 Projection mapping

The implementation of modern projection mapping involves the projection of light onto a surface, especially with the help of a camera as a sensor. Such surfaces are objects and walls or regular or irregular shapes, and often indoor environments with complex geometry. Projection mapping has also been applied to water[17, 23] and fog [21, 22] as display mediums.

### 4.2 Overview

The widespread use of projection mapping as a tool in digital art museums and live concerts recently have shown that projectors improved in resolution, dynamic range, frame rate, power consumption, colour gamut in recent years. The price for small companies to produce projection mapped artwork has also become cheaper, and projectors and software kits are available to consumers.

Projectors can be classified into two categories, those that have the goal of projecting undistorted assets onto (complex) geometry, and those that have the goal of controlling the color appearance of the projection. It is commonplace to have a mixture of both goals.

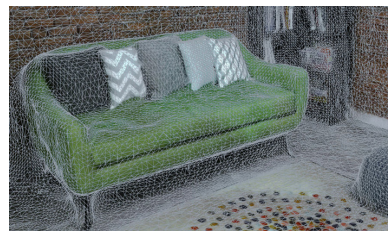

Figure 6: Mesh of watertight triangles to represent the geometry of the scene by surface [o]

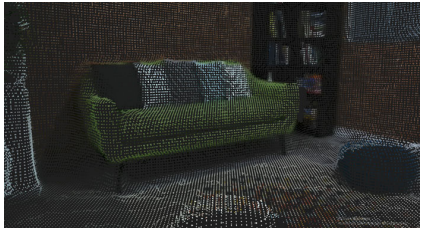

Figure 7: 3D point cloud of colored 3D points [o]

Full projection systems must go through calibrations, one of which is known as "geometric", which involves modelling the shape of the geometry of the projection surface through various techniques as well as physical parameters of the projectors and cameras. The other method is "photometric" which estimates internal color processing of the projectors and camera and attempts to reduce inaccuracies caused by the projection surface such as reflectance. The construction of the projected surface is either entirely known and modelled in computer aided software such as AutoCAD, or is reconstructed with a 3D reconstruction method such as 3D point clouds (see figure 7), the optimization of which is a research field of its own.

### 4.3 Calibrations

A visualization of the two main calibration methods are shown in Figure 2 [46]. In the geometric method, structured light patterns are captured by the camera to generate pixel correspondences between the projector-camera pair (procam). In the photometric method, color consistent projection is generated by projecting color patterns and sensed by cameras or colorimeters. The result of the combined two calibrations create an accurate image

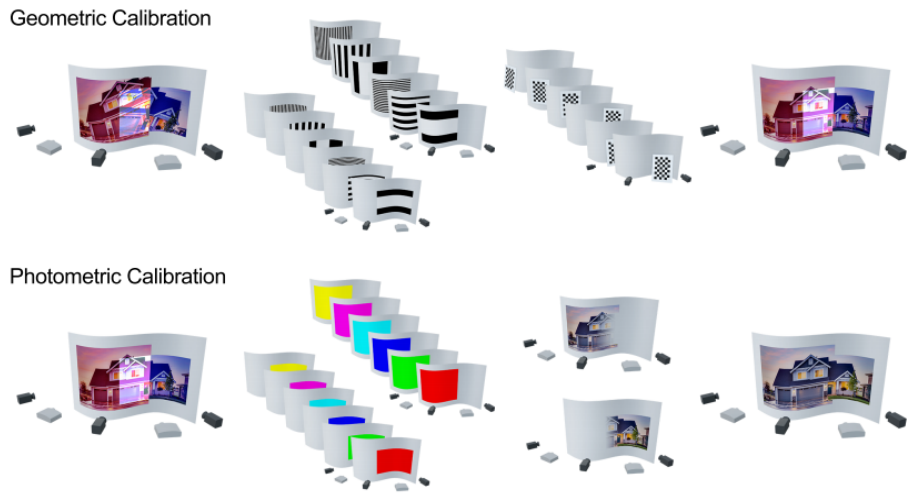

Figure 8: Combination of geometric and photometric calibration with procams.

The computer vision field in general requires accurate calibration of cameras and a method to register all input and output devices. This can be extended in the space of procam systems for interactive art experiences.

The most common method to calibrate cameras that is still applied in practice involves capturing multiple images of a marker board on a planar surface in order to estimate focal length, principal point, and error compensation parameters to model lens distortion [47]. To make calibration more quick and accessible to the general public such as artists, there have been numerous calibration methods. The next methods included treating the projector as the inverse of a camera using structured light patterns. A more recent paper proposed the projector to identify blob patterns it itself projected onto a planar surface that was out of focus for the projector [48]. The most recent techniques have involved developing an entire pipeline for multiple procam applications with semi-automated self-calibration. In a semi-automated method, the projector projects washing points to estimate internal and external parameters on a global scale but requires user guidance and three orthogonal planes.[49] Another proposed method requires the surface geometry of the space to be known beforehand [50]. Disney proposed the current most generic method for unknown surface geometries with no parameter tuning which frequently needs to be carried out by experts [51]. It is notable as it is the only calibration method that does not require initial guesses. It requires two or more cameras and self-adapts to outlier pixel correspondences. First, a pixel correspondence is generated using a projection of structured light, and pixel pairs considering outliers are matched. Then, it reconstructs the environment as a 3D point cloud and removes outliers after an iterative optimization function. As the next procams are added, it employs a fundamental matrix estimation of consecutive device calibration and surface reconstructions. Refer to figure 3 for a simplification of the process. Disney's method accurately calibrated 33 devices in a procam system in less than 45 minutes. Future work on the method proposes that the system will be developed to automatically suggest a number of cameras and optimal location positions for a given project.

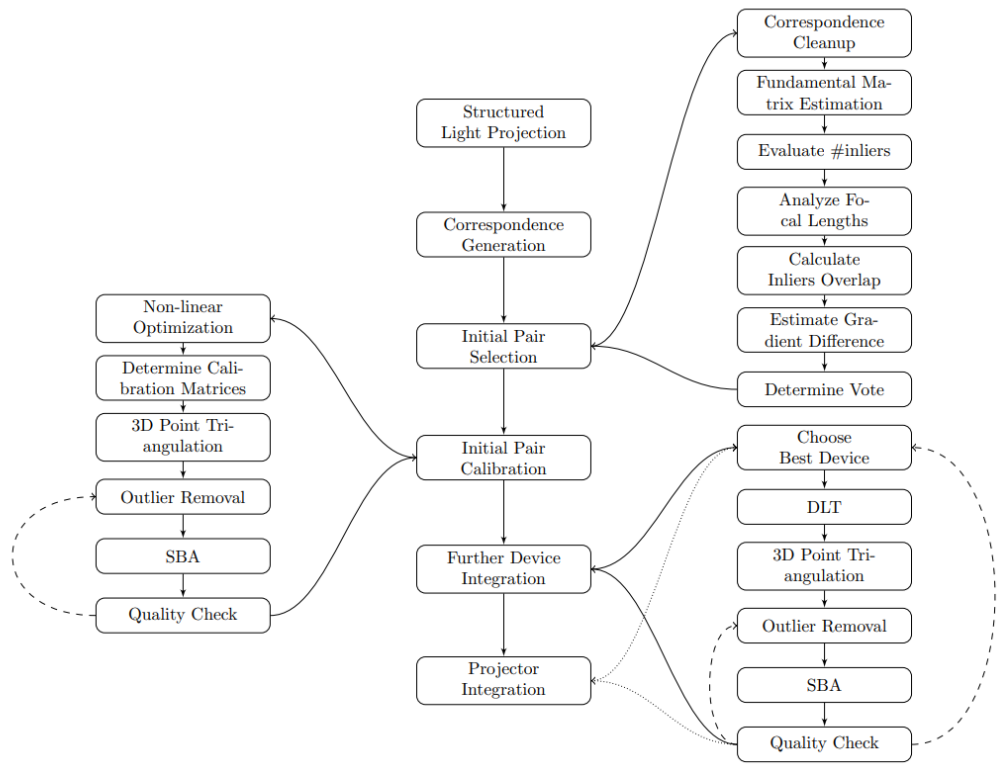

Figure 9: Disney's procam self-calibration algorithm

## 4.4 Rigid and non-rigid dynamic projection mapping

The field of projection mapping is developing methods for accurate projection in the case that a projected scene is able to move, or the projector or camera itself moves.

Rigid dynamic objects can be tracked with pose estimation of the projector. Obtaining a stable pose estimation can involve markers that are later diminished by the projector through radiometric compensation [52]. One such method achieved low latency since a 1000 Hz speed procam was used[53]. Pose estimation was also facilitated with a single IR camera [54] (figure 10) by solving a light transportation matrix obtained from depth of field relation between projectors and the object.

Deformable or "non-rigid" moving objects require higher accuracy methods for augmentation with projected light. A system was presented in 2017 using markerless human face tracking [53] with a latency of 10 ms. The system first uses a mesh (see figure 6) generated from facial expressions, then deforms it, applies a texture and uses an extended Kalman filter algorithm for motion prediction to correct the latency delay on the projected object. Depth sensors have been used to estimate pose with unaligned placement of depth cameras and projectors but with unsuitable latency for interactive displays [55]. Recently, high-speed projection mapping onto human arms has been performed within 10 ms of latency [83].

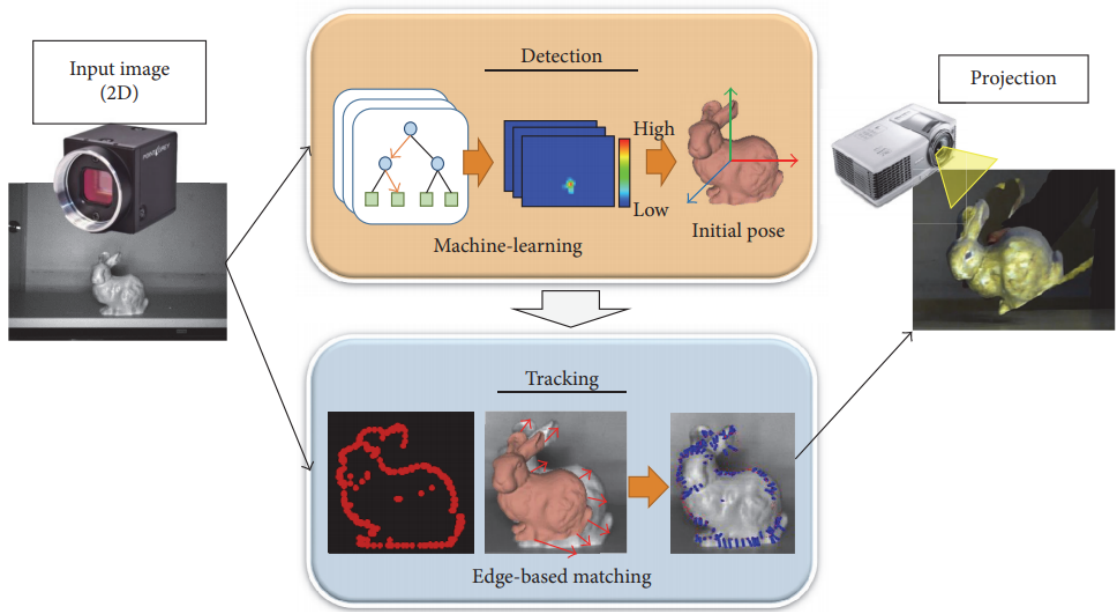

Figure 10: Pipeline of the single IR camera pose estimation and projection mapping system [f]

The conducted research used a parametric deformable model of human forearms, and a regression accuracy compensation method for the skin deformation based on texture mapping. The method however, requires predefined body shape parameters, but estimates joint position in real-time. The regression was trained on a dataset of forearms with markers. Possible applications of this implementation of projection mapping might include special effects wound make up, allowing an artist to draw on a participant in real-time, or showing options for tattoos.

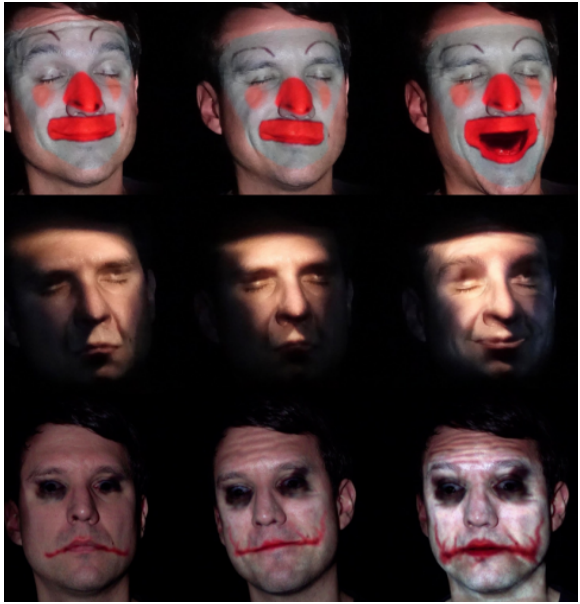

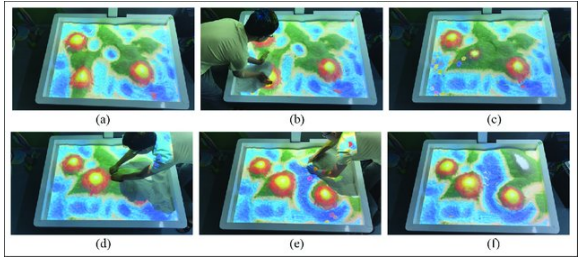

(Top) Figure 11: High-speed projection-mapped face paint deformed with facial expressions

(Bottom) Figure 12: Real-time rendering of "ocean, and volcanos" based depth of moved sand using a Kinect v2 RGB-D sensor and Epson CB-S04 projector

### 4.5 Hardware

Advances in hardware can rectify limitations in projection mapping algorithms.

Galvanoscopic laser projectors were calibrated in combination with procams since they provide a higher color gamut to display content [n]. Advances in hardware will attempt to accurately control and compensate for light emitted by projectors that unwantedly reflects from the target surface, building on algorithmic radiometric compensation techniques [56]. This might be through better production and therefore control of colors spatially and spectrally. Further research is needed to accurately project onto dynamic non-lambertian surfaces of complex textures.

### 4.6 Limitations

The high dynamic range (HDR) for a projection is currently only achieved in dark rooms or otherwise only for projection of static images.

High speed and low latency of projection systems used in various applications require improvement since touch interaction delay is common and disrupts immersion if feedback ie: by animation from a projection is greater than 6.04ms. Projectors that display 8-bit images running at 1 000 Hz can be improved to further reduce motion blur [57]. [58] The noticeable appearance of pixels still occurs when a visitor to a projection mapped exhibit sees the projected surface closely. Such a limitation is a technical challenge that needs to be solved. To increase focal depth and resolution, mitigation of contrast reduction caused by non-negative overlapping images and projected light in itself needs further investigation.

**\*mention cameras**

## 5.1 Implementation development tools and hardware

Display surfaces and hardware devices for extended reality including AR, VR and SAR maintain a large impact on the immersive aspect on users. Artistic effects and intentions are greatly limited or enhanced by choices in displays and interactivity methods due to technologies that have only seen the most widespread application in recent years. System development kits, tool kits, and other software has simplified the process of creating interactive art using cutting-edge technologies. In this section, an outline of displays, interactive devices, and toolkits currently in use are listed.

Toolkits such as WebXR now allow XR and AR content to be displayed on internet browsers. Projection mapping kits assist the projector calibration, 3D or 2D environment reconstruction for artists and projection mapping studios are readily available and have increased the spread of these applications.

Common Head-mounted displays are generally divided into fully immersive VR headset, and mixed-reality AR headsets. Hardware advances have sought to improve the tracking of the users hands, and gaze. Controllers compatible with the headsets commonly offer 6 degrees of freedom in movement to be tracked. A limitation of interactive art as applied to the "virtual experiences" field in general is the lack of haptic feedback. Currently proposed devices are discussed in section 5.2.

The tracking methods between devices differ, but use a modern pose estimation algorithm to identify gestures from the wearer. The most common approach when not dealing with groups of objects uses heatmaps on key-point detected joints of the wearer, applying graph theory to link the grouped joints to the image. A state-of-the-art study can be viewed here [59].

### 5.2 Toolkits

| Name | Category | Description/ Use case |
|------|----------|------------------------|
| WebXR | Web API | Display of VR and AR from any browser<br>api provided for collaborative virtual environments<br>support for accessing virtual reality (VR) and augmented reality (AR) devices, including sensors and head-mounted displays, on the Web |
| WebGL | Graphics library | |
| SparkAR | System Development Kit | Create AR experiences for Instagram |
| Unreal engine 4 | Game engine | Aid in game, animation, and graphics rendering through procedural animation "rigging" and physics engine. |
| Unity3D | Game engine | Aid in game, animation, and graphics rendering through procedural animation "rigging"and physics engine. |
| HeavyM[a7] | Projection mapping | Translate images and animations to projection mapped environments. Environment reconstruction information is accepted. |
| Madmapper | Projection mapping | Translate images and animations to projection mapped environments. Environment reconstruction information is accepted. |
| Notch | Projection mapping | Translate images and animations to projection mapped environments. Environment reconstruction information is accepted. |
| LucidWeb | Platform | Popular distribution platform for VR |
| Steam | Platform | Mainstream VR and AR game distribution |
| Gravity sketch | 3D modelling with VR software | -each stroke is a piece of 3D geometry, no drop down menus or hot keys. |

| | | | -V headset required<br>-ipad beta version |
|---|---|---|---|
| Flying shapes | 3D modelling with VR software | | -2D input to 3D graphic assets |

### 5.3 Head- Mounted displays (HMDs) display and controllers

| Name | Category | Display Type | Controller Depth of field | Field of view | Resolution per eye | Refresh rate |
|---|---|---|---|---|---|---|
| Magic leap | AR | AMOLED | 6 | 30x40 degrees | 1280x960 | 60Hz |
| Hololens2 | AR | AMOLED | 6 | 52 degrees | 2048 x 1080 | 120Hz |
| Oculus Rift S | VR | LCD | 6 | 90 degrees | 1280x1440 | 80Hz |
| HTC vive pro | VR | AMOLED | 6 | 110 degrees | 1440x1600 | 90Hz |
| HTC vive cosmos | VR | LCD | 6 | 110 degrees | 1440x1700 | 90Hz |
| Samsung HMD Odyssey+ | VR | AMOLED | 6 | 110 degrees | 1440x1600 | 90Hz |
| Valve index | VR | | 6 | 130 degrees | 1440x1600 | Option to select 80Hz<br>90Hz<br>120Hz<br>144Hz |

### 5.4 Head-Mounted displays (HMDs) tracking

| Name | Type | Method | Sensors |
|---|---|---|---|
| Oculus Rift S | Positional and AI assisted | Markerless, inside-out | 5 cameras |
| HTC vive pro | positional | Marker-based, inside-out | 2 IR emitting beacons |
| HTC vive cosmos | positional | Markerless, | 6 cameras |

| | | inside-out | |
|---|---|---|---|
| Samsung HMD Odyssey+ | positional | Markerless, inside-out | 2 cameras |
| Valve index | positional | Markerless, inside-out | 2 IR emitting beacons |
| Hololens2 | positional | Markerless, inside-out, -Hand and eye-tracking | 2 IR cameras 4 rgb cameras |

Inside-out VR device tracking systems estimate the user's position based on the data received from the environment. Marker-based tracking devices have mounted cameras that scan the surroundings to detect the markers placed at fixed points of an interaction area around the user processing and storing their spatial coordinates. In the markerless method, the position of the user is estimated using algorithms solely from camera and sensor input. [hmd]

### 5.4 Modern wireless haptic devices compared to "Wireality" [g8]

| | Wireality | VR Haptic Drones [26] | CLAW [10] | Normal Touch [5] | ElasticVR[51] | Wolverine [8] | DextrES [25] | PuPoP [50] | TORC [33] | Haptic Revolver [53] | Thor's Hammer [24] | Haptic Link [46] |
|---|---|---|---|---|---|---|---|---|---|---|---|---|
| Sensing Points | whole hand | whole hand | thumb + index | one finger | whole hand | thumb, index + middle fingers | thumb + index | whole hand | thumb, index + middle fingers | one finger | arm | arm |
| Actuator Type | ratchet gear with solenoid pawl + strings | drones | servo motor + voice coil actuator | servo motor | rotation motor + elastic band | one-way brake | electrostatic brake + piezoelectric actuator | pneumatic | vibrotactile motor + voice coil actuator | DC motor + wheel | motor + propeller | linear actuator + ball-and-socket |
| Haptic feedback | collision | collision | collision + grip + texture | collision + compliance | impact + compliance | grip | grip + compliance | grip + compliance | texture + compliance | texture | tug | bimanual grip |
| Max. Force per Finger | 183N | n/a | 30N | n/a | 14N | 106N | 20N | n/a | n/a | 3.35N | 4N | 80N |
| Peak Power Consumption | 2.19W | 20.40W | 5W | n/a | n/a | 2.89W | <0.12W | 24W | n/a | 2.50W | 204.70W | 7.20W |
| Weight on the Hand | 11g | 0g | 420g | 150g | 150g | 55g | 16g | n/a | n/a | 237g | 692g | >651g |
| Cost | $35 | $30 | >$150 | n/a | n/a | $40 | n/a | n/a | <$90 | n/a | n/a | n/a |

Table 1. A comparison of prior hand-centric haptic systems and Wireality.

### 5.5 RGB-D Cameras [rgb]

| Name | Type | Field of view | Frame rate | 3D resolution | Depth range |
|---|---|---|---|---|---|
| Microsoft kinect 2.0 | Time of flight | 70 degree horizontal (H), 60 degree vertical (V) | 30 fps | 512 x 424 | 0.5 to 4.5 m |
| Asus XtionPro Live | Structured light | 58 degree H, 45 degree V | 30 fps | 640 x 480 | 0.8 to 3.5m |

| Intel RealSens Camera D435i | Active IR Stereo using Global Shutter Sensors and IMU | 85.2° x 58° (+/-3°) | 30 fps at max depth resolution; up to 90fps at lower depth resolution | 1280 x 720 max | 0.105 to 10 m |
|---|---|---|---|---|---|

### 5.6 Haptics

Haptic devices used to provide realistic feedback in virtual and mixed reality environments in the context of interactive art displays and experience. Recently, a soft texture database was added to OpenHaptics database with Phantom Omni recorded textures [g1]. Wireless gloves without finger tracking devices such as a Leap Motion continue to be developed [g2]. In one implementation case, a hand is tracked by an IR camera which detects three infrared LEDs attached to the 3D printed base on the glove. Object elasticity has been explored through large or global deformation with a FEM-based approach [g3] and otherwise [g6]. Other wireless devices are outlined. As well, a new concept of distributed haptic display for realistic interaction with a given virtual object of with shape complexity by a collaborative robot and shape display end-effector has been developed [g4]. A palette-type controller is another new device presented in recent times [g5]. Magnets fields as haptic feedback devices were studied in the demonstration of tactile interactions [g7]. Modern wireless wire haptic devices (shows all modern drones, etc) are compared to a novel hand-attached-to-string approach in table 5.4 [g8].

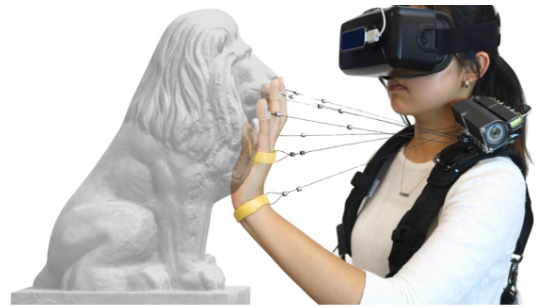

Figure 12: Wireality [g8]

### 6.1 Artistic Effects and Animation

Recent research innovations for artists will increase the types of interactive art experiences available to the public. In this section, we present imagery, projection and graphics effects demonstrated in lab settings that have a high chance of adaptation by artists. Finally, we discuss current techniques in procedural animation employing machine learning techniques.

### 6.2 Novel artistic effects

Artificial digital colorization of 3D real-world objects [80] has been developed for the purpose of using projection mapping to map realistic colours onto ancient artifacts. An archaeologist or restoration expert can utilize the proposed "proxy painting" method using a paintbrush by painting paint onto a smaller 3D printed model of the targeted artefact, which is tracked, rendered in real-time using Unreal Engine 4, and projection mapped onto the targeted real object. This implementation avoids occluders resulting from implementations that attempted to allow a tracked device near target`s projection surface and replaces the need for haptic feedback and capture 3D gestures from the paintbrush itself [81]. No method was applicable to complex shaped three-dimensional surfaces due to varied lighting and self-shadowing without tracking a brush with size degrees of freedom. This method translates the paint into texture space and is an available plugin for Unreal Engine 4.

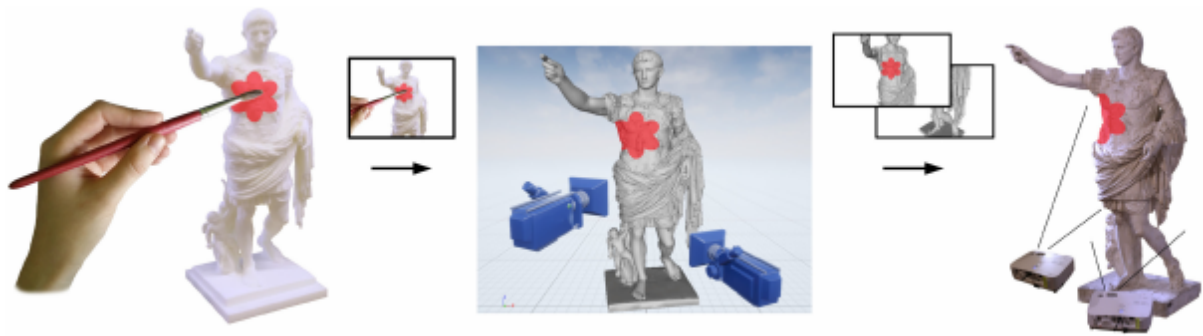

Figure 13: The painting of the colored proxy object is captured per frame and projected onto the target object

A system to enable dynamic control of a projector area in high resolution using a static image is proposed with proper geometrical alignment, overcoming the trade off between the angle of projected media and the resolution [82]. The research introduced a laser pointer to be tracked by a camera sensor, and a high-speed optical axis controller composed of shift lenses and rotational mirrors with a total system latency of 3ms and 1000 fps image projection. In art, this concept can be developed to use eye tracking to interactively unveil artwork based on the gaze of a viewer and perhaps on complex geometries [video10].

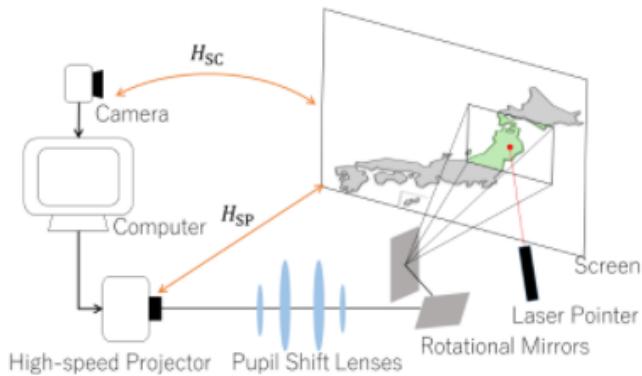

Figure: Projection mosaicing system with camera sensor, projector and optical axis controller

A generic framework for modelling the interaction of projected light and a pattern surface "Paxel", has applications for color-changing effects. In the experiment, a 365nm UV light was projected onto a rgd fluorescent pattern inked surface and white surface. Specific colour changes are created for example, in figure 9 a) highly saturated red and blue colors are projected onto a white surface, and results in a red surface appearance, while projecting it on a pattern surface, the red light is absorbed by the black ink and the surface appears blue. In b) projecting a red color on red absorbing colorants while projecting a low saturated blue on blue reflecting colourants. When projecting onto a white surface, the red remains and overtakes the blue, whereas projecting onto a pattern surface absorbs the red, and the pattern surface appears more blue. In c) projecting a white light onto a colored pattern surface tricks the eye into averaging rgb intensities to form color.

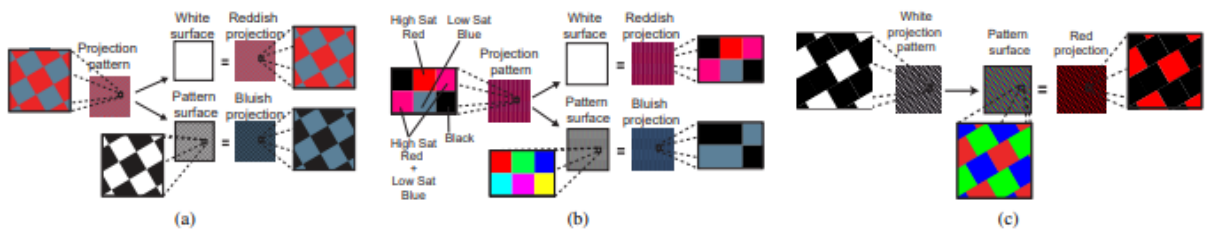

Figure 14: a) black-white pattern surface, b) colored pattern surface, c) monochrome-chrome projection

The novel method of color change obtained an average error $\Delta E^*_{00}$ between the mean and measured color change for the prediction through the framework. Paxel can be applied to illusionary-like performance art, since UV light can be projected onto fog with high quality, unlike other methods. The color-changing effect can be used extensively in art exhibitions such as applying the concept of stenography. For example, two pages put together will form an image that was previously not able to be seen. In this case it is the coding of which pattern the surface uses, how many colorants there are, and how many clusters, and this can be triggered or given to certain participants. In the future, the work can be extended to different color spaces. Figure 10 explains explored practical applications of the framework.

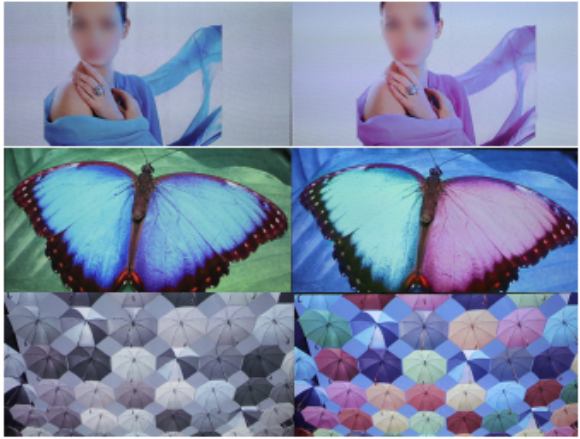
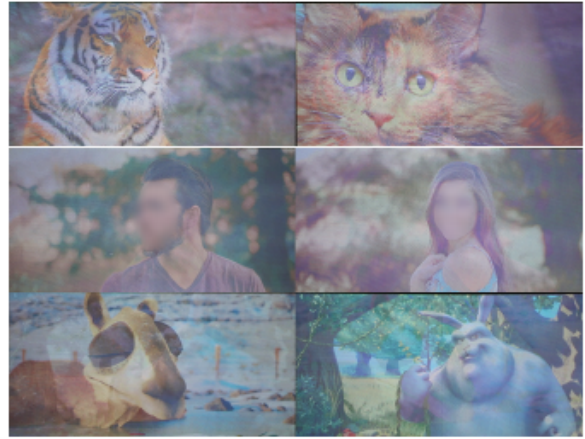

Figure 15: Paxel applied for color-changing effect

Limitations in the framework exist, such as the resolution, and the fact that the display only covers a limited color gamut in CIELAB colorspace. The framework still needs to be optimized to process images in real-time for live-content.

### 6.2 Neural rendering

The field of graphics rendering has improved by neural network rendering methods in recent years and a popular method "DeepFakes" have been introduced which utilize General Adversarial Networks (GANs). The most successful method used in images is conditional GANs [92]. It is most commonly used to swap the faces of people in videos and images, but has applications spanning as far as indoor floor plan generation. This method can be used to reduce production costs for films by paying a studio to train a network to replace an actor's face in film footage a production studio already owns with a target actor and therefore create a new video by "face-swapping" the old actor with a new actor, in an old movie, or vice-versa, without the need to hire an actor. An amateur film made in this way has already been released [93]. It does not require extensive training data. An application drawn from a GAN approach could take the form of digital art museums such as "Borderless" enabling digital twins of museum visitors rendered in a style of their choice, or online platforms that render images of your choice into the style of a famous artist as shown below.

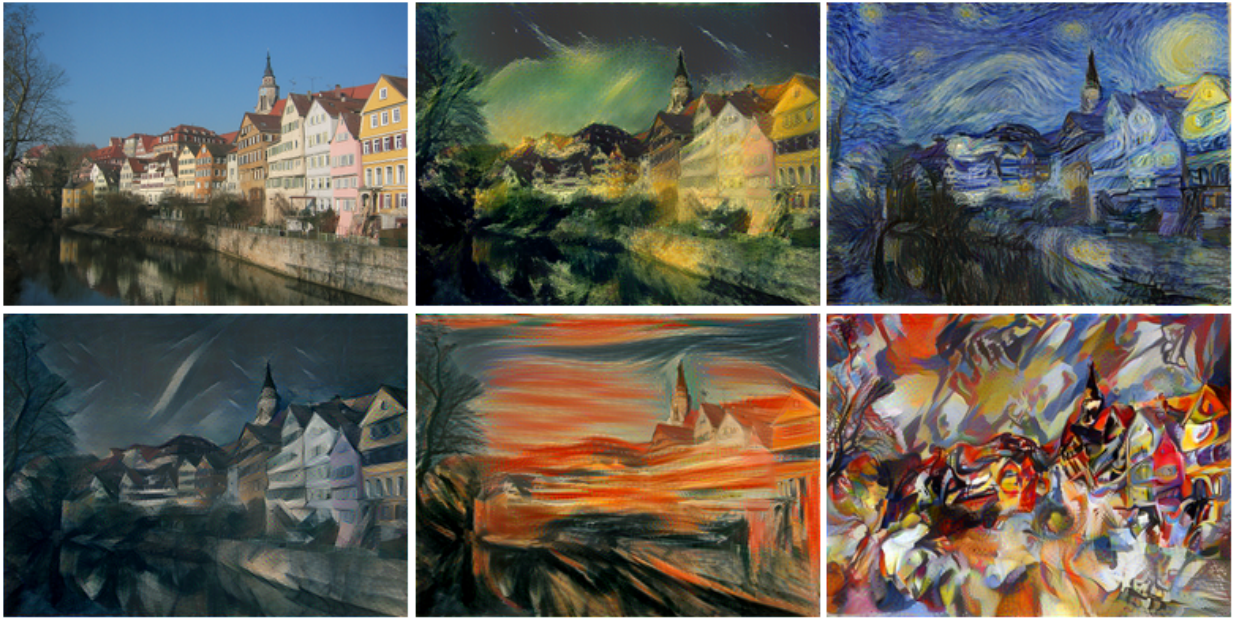

Figure 16: DeepFake regeneration of a street in the style of Van Gogh's starry night (top right) and other famous artists

This idea of face-swapping has recently been improved to preserve lighting and contrast after the swap and to improve inpainting. [94] Disney created this technique using variational autoencoders. An encoder distils a high dimensional input such as an image source into lower dimensional vector space (latent space) and a decoder reconstructs the source content or target content given the aforementioned vector. Usually this framework is trained end-to-end until consistency is reached. The latent space can be controlled to affect the output such as interpolating between the two inputs, and generating new content. The method first detects a face, and localizes facial landmarks, the image is normalized to 1024x1024 resolution, and fed to the network, saving the output of the s-th decoder. Then, the image's normalization is reversed, and the save parameters are poisson blended with a compositing method. It is the most accurate face-swapping method known to date, but is limited in proper face-swaps of profile views and certain expressions such as squinting which do not contain many facial landmarks. This can be solved by a smaller gap in training data. All other comparable methods also do not change the face shape, or people wearing glasses and this is a topic for further research in the field. It is the first method that achieves convincing face swaps on high resolution (1024x1024) media.

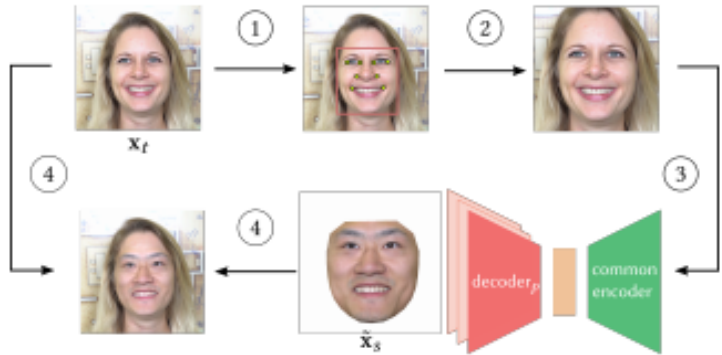

**Figure 2:** *A schematic of the full pipeline for swapping a source face of identity s onto a person t ≠ s. In steps (1) and (2) we pre-process the input by cropping and normalizing the face. In step (3) the pre-processed image is fed into the common encoder and decoded with corresponding decoder $D_s$. In (4) we use our multi-band blending to swap the target with the source face.*

Figure 17: Disney's face-swapping pipeline

Another application of neural rendering enables a character in a video game to be automatically properly animated, for example, to run over complex terrain [95]. In interactive art, the method could be leveraged to allow users to create game characters that adventure in a virtual world. It can also be applied to improve realism of movement in virtual environments.

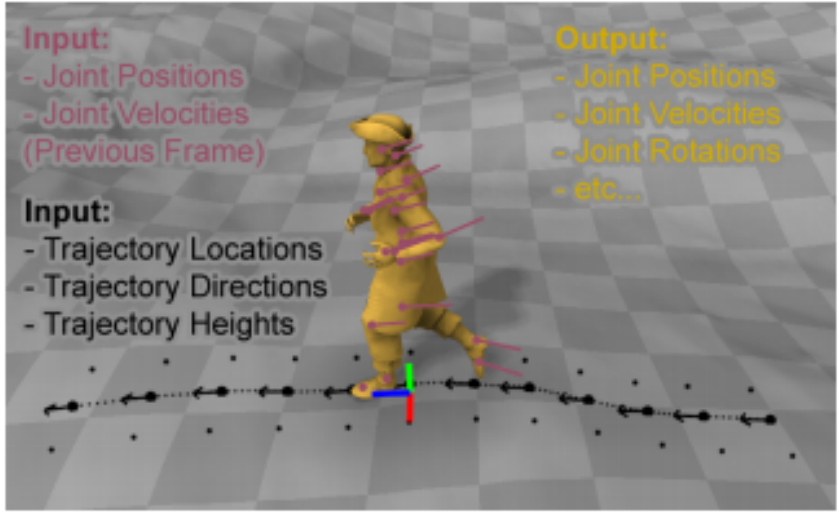

Figure 18: The terrain animation system presented by [95]

A broad view of current innovations in neural rendering include novel-view synthesis and free viewpoint videos where objects can be viewed from any angle, realistic relighting, reenactment of captured body and face data to representative digital avatars, and semantic image editing. Digital avatars have been created through machine learning techniques without manual modeling. The reader is referred to a state-of-the-art report that details this topic [96].

**7 Conclusion**

This state of the art review Explored emerging applications of interactive art experiences and novel innovations in animation technology through the outlining of research in the past 3 years. The interpretation of technology and art through technology will continue to grow as XR and projection mapping continue to expand into mainstream markets.

[37] Black, J., Bradshaw, J., Cokas, C., Ham, K., McNair, E., Rooney, B., & Swartz, J. (2020). ESPN VR Batting Cage. In *ACM SIGGRAPH 2020 Immersive Pavilion*. Association for Computing Machinery.

[38] Oculus Blog. (2020, September 1). *Venice international film festival brings VR selection to oculus*. Oculus | VR Headsets & Equipment. https://www.oculus.com/blog/venice-international-film-festival-brings-vr-selection-to-oculus/

[39] Cutler, L., Tucker, A., Schiewe, R., Fischer, J., Dirksen, N., & Darnell, E. (2020). Authoring Interactive VR Narratives on Baba Yaga and Bonfire. In *Special Interest Group on Computer Graphics and Interactive Techniques Conference Talks*. Association for Computing Machinery.

[40] Ligthart, M., Neerincx, M., & Hindriks, K. (2020). Design Patterns for an Interactive Storytelling Robot to Support Children's Engagement and Agency. In *Proceedings of the 2020 ACM/IEEE International Conference on Human-Robot Interaction* (pp. 409–418). Association for Computing Machinery.

[41] Vilk, J., & Fitter, N. (2020). Comedians in Cafes Getting Data: Evaluating Timing and Adaptivity in Real-World Robot Comedy Performance. In *Proceedings of the 2020 ACM/IEEE International Conference on Human-Robot Interaction* (pp. 223–231). Association for Computing Machinery.

[42] http://www.wired.com/2017/02/lady-gaga-halfime-show-drones/

[43] Tsuru Robotics (2018, March 29). First in the World Graffiti Drone (Part 1). Retrieved at htps://diydrones.com/profiles/blogs/first-in-the-world-graffitidrone-part-1, 2018, December 15.

[44] Andrade, P. (2019). Mobile Social Hybrids and Drone Art. In Proceedings of the 9th International Conference on Digital and Interactive Arts. Association for Computing Machinery. 4-5

[46] https://s3-us-west-1.amazonaws.com/disneyresearch/wp-content/uploads/20180406102342/Recent-Advances-in-Projection-Mapping-Algorithms-Hardware-and-Applications-Paper.pdf?fbclid=IwAR1_wlfyIFhmEXIZ5j69J7Q0OM03XQqKQdohlDaU0s4_xC928Ylf6MbemJ8#page=3&zoom=100,68,78

[r] https://studios.disneyresearch.com/2018/10/02/paintcopter-an-autonomous-uav-for-spray-painting-on-3d-surfaces/

[t] https://www.oculus.com/experiences/quest/1891638294265934/?locale=fr_FR

[u] Fazio, S., & Turner, J. (2020). Bringing Empty Rooms to Life for Casual Visitors Using an AR Adventure Game: Skullduggery at Old Government House *J. Comput. Cult. Herit., 13*(4).

[v] Goto, S. (2019). GravityZERO, an Installation Work for Virtual Environment. In *Proceedings of the 6th International Conference on Movement*